\documentclass[12pt]{article}
     
\usepackage{latexsym}
\newcommand{\eqn}[1]{(\ref{#1})}
\newcommand{\ft}[2]{{\textstyle\frac{#1}{#2}}}
\newsavebox{\uuunit}
\sbox{\uuunit}
    {\setlength{\unitlength}{0.825em}
     \begin{picture}(0.6,0.7)
        \thinlines
        \put(0,0){\line(1,0){0.5}}
        \put(0.15,0){\line(0,1){0.7}}
        \put(0.35,0){\line(0,1){0.8}}
       \multiput(0.3,0.8)(-0.04,-0.02){12}{\rule{0.5pt}{0.5pt}}
     \end {picture}}

\def\be{\begin{equation}}
\def\ee{\end{equation}}
\def\ba{\begin{array}}
\def\ea{\end{array}}
\def\bea{\begin{eqnarray}}
\def\eea{\end{eqnarray}}
\def\bd{\begin{document}}
\def\ed{\end{document}}
\def\ts{\textstyle}
\let\la=\label
\let\bm=\bibitem
\def\nn{\nonumber}
\def\qq{\quad\quad}

\def\ft#1#2{{\textstyle{{\scriptstyle #1}\over {\scriptstyle #2}}}}
\def\fft#1#2{{#1 \over #2}}

\def\sst#1{{\scriptscriptstyle #1}}
\def\oneone{\rlap 1\mkern4mu{\rm l}}

\newcommand{\eq}[1]{Eq.~(\ref{#1})}
\newcommand{\sect}[1]{Sec.~(\ref{#1})}

\newcommand{\eqs}[2]{Eqs.~(\ref{#1})-(\ref{#2})}

\def\Hat#1{\widehat{#1}}

\def\a{\alpha}
\def\b{\beta}
\def\c{\gamma}
\def\C{\Gamma}
\def\d{\delta}
\def\D{\Delta}
\def\e{\epsilon}
\def\vare{\varepsilon}
\def\eb{{\bar\epsilon}}
\def\F{\Phi}
\def\vf{\varphi}
\def\k{\kappa}
\def\l{\lambda}
\def\L{\Lambda}
\def\m{\mu}
\def\n{\nu}
\def\r{\rho}
\def\s{\sigma}
\def\S{\Sigma}
\def\th{\theta}
\def\Th{\Theta}
\def\o{\omega}
\def\O{\Omega}
\def\g{\gamma}
\def\cD{{\cal D}}
\def\cF{{\cal F}}
\def\cM{{\cal M}}
\def\cR{{\cal R}}
\def\cDs{\not\!\!{\cal D}}
\def\lra{\leftrightarrow}
\newcommand{\dsl}{\not\!\partial}
\def\Dsl{\not\!\!{\cal D}}
\newcommand{\bslash}{\not\! b}
\newcommand{\omslash}{\not\! \omega}
\newcommand{\vslash}{\not\! V}
\newcommand{\dr}{\raise.3ex\hbox{$\stackrel{\leftarrow}{\partial }$}{}}
\newcommand{\delr}{\raise.3ex\hbox{$\stackrel{\leftarrow}{\delta }$}{}}
\newcommand{\eentwee}{-1 \leftrightarrow 2}
\newfont{\twolineletters}{msbm10}
\newcommand{\Zzbar}{\mbox{\twolineletters Z}}
\renewcommand{\theequation}{\thesection.\arabic{equation}}
\csname @addtoreset\endcsname{equation}{section}

\begin{document}
\begin{titlepage}
\begin{raggedleft}
KUL-TF/99-31\\
{\tt hep-th/9910175}\\[.5cm]
October 1999\\
\end{raggedleft}
\ \vskip2cm
\begin{center}
{\bf \Large An action for the $(2,0)$ self-dual tensor multiplet in a conformal supergravity background}
\vskip 10.3mm
{\bf Kor~Van Hoof}
\vskip1.5cm
Instituut voor Theoretische Fysica, Katholieke
Universiteit Leuven,\\ Celestijnenlaan 200D, B-3001 Leuven, Belgium\\
\end{center}
\vfill
\begin{center}
{\bf Abstract}
\end{center}
\begin{quote}
We present the action for a self-dual tensor in six dimensions, coupled to a $(2,0)$ conformal supergravity background. This action gives rise to the expected equations of motion. An alternative look upon one of the gauge symmetries clarifies its role in the supersymmetry transformation rules and the realization of the algebra.
\end{quote}
\end{titlepage}
\section{Introduction}

The description of chiral bosons, antisymmetric $2p$-forms in $4p+2$ dimensions with (anti)self-dual field strengths, in terms of an action has been a longstanding problem. This was caused by the first order self-duality condition. In $6$ dimensions, the chiral boson is a (anti)self-dual two-tensor. By giving up explicit Lorentz invariance, it became possible to construct a new class of actions \cite{HT,Schwarz}. In \cite{Schwarz}, six-dimensional actions were constructed with manifestly five-dimensional Lorentz invariance. A non-trivial check was needed to prove six-dimensional covariance. 
\par The construction of a manifestly Lorentz invariant action has been established by introducing {\sl one} auxiliary scalar field and two extra gauge symmetries in \cite{PST1}. This auxiliary field appears non-polynomially in the action and in the two new local gauge symmetries. These bosonic symmetries make it possible to gauge away the new scalar field and reduce the number of degrees of freedom of the tensor. The gauge fixing of these symmetries in the bosonic model is studied in \cite{vdbvh}.
\par In \cite{lechner} was proven that this Lorentz invariant action with a suitable gauge fixing (choosing a fixed direction for the derivative of the scalar) gives rise to the conjectured Feynman rules and the gravitational anomalies as in \cite{Witten}.
\par Self-dual tensors also appear in a supersymmetric context. Depending on the amount of chiral supersymmetry in six dimensions, the supersymmetric tensor multiplet contains a (anti)self-dual tensor, 2 or 4 chiral spinors and one or five scalars. In \cite{ckvp} is proven that the action for the rigid $(2,0)$ tensor multiplet is invariant under rigid superconformal symmetry. The geometrical superspace treatment of the self-dual tensor multiplet is developed in \cite{grojean}.
\par The field equations for the $M5$-brane \cite{pst2, pst3, HSW} also contain a self-dual tensor. The worldvolume action is constructed in \cite{pst2}. The WZ-term and the relation to the rigid superconformal model appeared in \cite{ckvp}. Up to now, a model for interacting self-dual tensor multiplets is lacking. In \cite{bhs} is proven that this will not be a local interacting quantum field theory. The superconformal theory for a stack of $n$ coinciding $M5$-branes would be relevant for the adS$_7$/CFT$_6$ version of the Maldacena conjecture. Some partial results can be found for instance in \cite{leighrozali}.
\par Self-dual tensors appear naturally in string theory compactifications, for instance compactifications of type $I$ string theory on $K_3$ \cite{sagnotti}, of the heterotic string on $K_3$ \cite{walton} and of $M$-theory on $K_3 \times S^1/\Zzbar_2$ \cite{sen} in the $(1,0)$ context and  compactifications of type $IIB$ on $K_3$ \cite{seiberg} with $(2,0)$ supersymmetry in six dimensions. The (anti)self-dual tensors appear in two ways in gravitational theories. First, they are one of the components of the gravitational multiplet in Poincar\'e \cite{6dsugra, romans} and conformal \cite{BSVP1, BSVP} chiral supergravities in six dimensions. The rigidly supersymmetric tensor multiplets can also be realized as matter multiplets in a supergravity background. In \cite{DLT}, the Poincar\'e actions for the multiplets with self-dual tensors were constructed. The most general coupling of matter multiplets in chiral theories with one supersymmetry can be found in \cite{nishsez}. These chiral theories are anomalous. The $(2,0)$-theory is anomaly-free when 21 tensor multiplets are coupled to supergravity \cite{townsend}.

\par By using superconformal techniques \cite{sctc}, the field equations for the self-dual tensor multiplet coupled to a  $(1,0)$ and a $(2,0)$ conformal supergravity background were derived in \cite{BSVP1} and \cite{BSVP}. In \cite{BSVP}, the gauge-fixing of the relevant superconformal symmetries to end up with super-Poincar\'e symmetry, was done and the description to go from the field equations of the tensor multiplet with $(2,0)$ to those with $(1,0)$ superconformal symmetry was given. 
\par The aim of this paper is to give the {\it action} for the self-dual tensor multiplet, coupled to conformal $(2,0)$ supergravity. The consistency of the action is checked by the derivation of the field equations of \cite{BSVP}. We give the  transformation rules of the tensor multiplet that do not use the self-duality condition. These transformation rules are understood better by looking upon the first new symmetry as a gauge symmetry with the derivative of the auxiliary scalar as its gauge field. We also indicate how the algebra is changed when these transformation rules for the tensor multiplet are used and the new gauge symmetries are introduced. The transformation rules of the equations of motion are given.
\par The paper is organized as follows. In section~\ref{s:suconfmp}, the field content of the Weyl multiplet is given. The $(2,0)$ tensor multiplet, its  superconformal transformation rules and its gauge transformation with respect to the new gauge symmetries are studied. Therefore, the auxiliary scalar field $a$ is introduced. All the three gauge symmetries play a specific role in the algebra. In section~\ref{s:action}, the action and a sketch of the proof of its invariance under gauge and superconformal symmetry is given. This action gives rise to the field equations of \cite{BSVP}. Their (special) supersymmetry transformation rules are given. The last section contains the conclusions.

\section{$(2,0)$ Superconformal multiplets} \label{s:suconfmp}

In this section, the superconformal gravitational multiplet and the self-dual tensor multiplet are introduced\footnote{The notations, conventions and the transformation rules for the Weyl multiplet of \cite{BSVP} are used.}. The new gauge symmetries, essential to write down an action, are defined. The role of the new symmetries in the transformation rules and the algebra is clarified.

\par The $(2,0)$ superconformal gravitational multiplet is a representation of $OSp(8^*|4)$ \cite{ckvp}. The usual superconformal tensor calculus approach is followed to construct the Weyl multiplet: introducing gauge fields for all the superconformal symmetries, constructing curvatures, imposing conventional constraints and  introducing matter fields. The Weyl multiplet then contains gauge fields and matter fields. $e_\m{}^a$, $\psi_\m^i$, $V_\m^{ij}$ and $b_\m$ are gauge fields for general coordinate transformations, supersymmetry, $USp(4)$ $R$-symmetry and dilatations. The gauge fields $\omega_\mu{}^{ab}$ for Lorentz rotations, $f_\m{}^a$ for special conformal transformations and $\phi_\mu^i$ for special supersymmetry are composite fields by the three conventional constraints. Together with the matter fields, five antiself-dual tensors $T^{ij}_{abc}$, sixteen fermions in  $\chi_{jk}^i$ and fourteen scalars in $D^{ij,kl}$, they give rise to $128+128$ degrees of freedom. 

\par There is only one matter multiplet in the chiral $(2,0)$ theory: a self-dual tensor multiplet\footnote{The antiself-dual tensor multiplet is a representation of the $(0,2)$ antichiral algebra.}. 
It contains a self-dual tensor $B_{\mu\nu}$ with three degrees of freedom, a set of four symplectic Majorana-Weyl fermions $\psi^i$ and five scalars $\phi_{ij}$. Some properties of these fields are summarized in table~\ref{tbl:tensormultiplet}. In the definition of the fermions and the scalars, we made use of $\Omega_{ij}$, the $USp(4)$-metric. $USp(4)$-indices are moved up and down using this metric as follows:
\be
\l^i = \Omega^{ij}\l_j\, , \qquad \l_i = \Omega_{ji}\l^j\, .
\ee

\begin{table}[htb]
\begin{center}
\begin{tabular}{||c|c|c|c|c||}
\hline
Field & Type & Restrictions & {\rm USp(4)} & w\\
\hline
&&&&\\
$B_{\mu\nu}$   & boson   &real antisymmetric tensor gauge field&1&0\\
&&&&\\
$\psi^i$   & fermion &$\gamma_7\psi^i = -\psi^i$&4& $\ft 52$\\
&&&&\\
$\phi^{ij}$& boson   &$\phi^{ij} = -\phi^{ji} \ \ \ \
\Omega_{ij}\phi^{ij} = 0$&5&2\\
&&&&\\
$a$        & boson   & & 1& 0 \\
&&&&\\
\hline
\end{tabular}
\caption{ The fields of the $(2,0)$ tensor multiplet and the auxiliary scalar $a$ with  the various algebraic restrictions on the fields,
their USp(4) representations assignments and the Weyl weights $w$.\label{tbl:tensormultiplet}}
\end{center}
\end{table} 
We start from the following on-shell (special) supersymmetry transformation rules \cite{BSVP}:
\bea\label{susyBmn}
\d B_{\m\n} &=& - {\bar\e}^i\gamma_{\m\n}\psi_i +{\bar\e}^i
\c_{[\m}\psi_{\n]}^j\phi_{ij}\ , \\
\label{susy1psi}
\d \psi^i & = & \ft 1{48} H^{+}_{\m\n\r}\gamma^{\m\n\r} \e^i
+{\ts{1\over 4}}\not\!\! \cD \phi^{ij}
\e_j - \phi^{ij}\eta_j\ , \\
\label{susyphi}
\d \phi^{ij} &=& -4
{\bar\e}^{[i}\psi^{j]} -\Omega^{ij}{\bar\e}^k\psi_k .
\eea
In (\ref{susy1psi}), the first term contains the self-dual part of the covariant field strength for the tensor:
\begin{equation} \label{covH}
H_{\mu\nu\rho} \equiv 3\partial_{[\mu} B_{\nu\rho]} + 3\bar\psi_{[\mu}^i\gamma_{\nu\rho]}\psi_i -\ft 32 \bar \psi_{[\mu}^i\gamma_\nu\psi_{\rho ]}^j\phi_{ij} 
- \ft 12 \phi_{ij}T^{ij}_{\mu\nu\rho} \, ,
\end{equation}
where
\begin{equation}
H^{\pm}_{\mu\nu\rho} = \ft 12 \left(H_{\mu\nu\rho} \pm \ft 1{3!} \varepsilon_{\mu\nu\rho\sigma\tau\phi} H^{\sigma\tau\phi}
\right)\, .
\end{equation}
The definition of the covariant field strength in \eqn{covH} implies a choice for the matter term, which was not included in its definition in \cite{BSVP}.  
To realize the algebra with the supersymmetry transformation rules \eqn{susyBmn}-\eqn{susyphi}, the self-duality condition,
\be\label{selfdual}
H^-_{\m\n\r} = 0\, ,
\ee 
must be used. In section \ref{s:action} is explained how this condition is found from the action (\ref{action}).

\par The action for the self-dual tensor can be constructed by making use of one auxiliary scalar $a$. This scalar does not transform under supersymmetry:
\be
\d_Q a = 0 \, .
\ee
Its derivative is $u_\m = \partial_\m a$. It always appears in the action in the following way:
\begin{equation}
v_\m \equiv \frac{u_\mu}{\sqrt{u_\nu g^{\nu\rho}u_\rho}}\, .
\end{equation}
The contraction of the ((anti)self-dual) field strength with $v_\m$ is defined as
\be
H_{\mu\nu} \equiv H_{\mu\nu\rho} v^{\rho}\, , \qquad H^\pm_{\mu\nu} \equiv H^\pm_{\mu\nu\rho}v^\rho\, .
\ee
There are three local bosonic gauge symmetries that act on the tensor and the field $a$.
The first one is the usual reducible gauge transformation for a antisymmetric tensor:
\be\label{symm1}
\d_I B_{\m\n} = 2\partial_{[\m}\a_{\n ]}; \qquad \d_I a = 0\, .
\ee
This gauge symmetry has a reducible component:
\be
\a_\m = \partial_\m \a\, .
\ee
The two other gauge symmetries act in the following way on $B_{\m\n}$ and $a$:
\bea\label{symm2}
\d_{II} B_{\m\n} & =  & 2 H^-_{\m\n}\frac{\phi}{\sqrt{u^2}} \qquad \d_{II} a = \phi \, ,\\ 
\d_{III} B_{\m\n} & = & u_{[\m}\Lambda_{\n ]}\, .\label{symm3}
\end{eqnarray}
All the other fields are inert under each of these three symmetries. 
The $H^-_{\m\n}$ in $\d_{II}$ contains the covariant field strength (\ref{covH}).  Also symmetry $III$ has a zero mode: $\Lambda_\m = u_\m \Lambda$. A last reducible gauge transformation is formed by a combination of $I$ and $III$: $\a_\mu = u_\mu \Psi
$ and $\Lambda_\mu = 2 \partial_\mu \Psi$.\\
The divergence of the auxiliary field can be considered as the gauge field of the second symmetry, since $u_\m$ transforms into the derivative of the parameter:
\be\label{gauge}
\d_{II} u_\m = \partial_\m \phi\, .
\ee

\par The commutators of these gauge symmetries are:
\bea
\left[ \d_{II}(\phi_2), \d_{II}(\phi_1) \right]  
& = & \d_{III} (4 \frac{H^-_{\m\n}}{(u^2)^{3/2}}  
(\phi_1 \partial^\n \phi_2 - \phi_2 \partial^\n \phi_1))\, , \\
\left[ \d_{II}(\phi), \d_{III}(\Lambda_\m) \right] & = & \d_I 
(\ft 12 \Lambda_\m \phi)
+ \d_{III}(2\partial_{[\m} \Lambda_{\n]}\cdot u^\n {{\phi}\over{u^2}})\, . 
\eea
Their role in the gauge-fixing of the bosonic symmetries can be found in \cite{vdbvh}.

\par Making use of these gauge symmetries, it is possible to write down other supersymmetry transformation rules for the spinors that realize the algebra without using the self-duality condition (\ref{selfdual}). This comes already closer to an off-shell realization of the algebra, since the self-duality condition \eqn{selfdual} must not be used any more.
Therefore, we will replace the self-dual component of the field strength in (\ref{susy1psi}) by another self-dual tensor (as in \cite{DLT}), such that it gives the old transformation rule upon imposing the self-duality condition (\ref{selfdual}):
\begin{equation}
h_{\mu\nu\rho}^+ \equiv \ft14 H_{\mu\nu\rho}
-\ft32 v_{[\mu}H^-_{\nu\rho]}\, .
\label{defh}
\end{equation}
We can rewrite the terms with $v_\m$ in $h^+_{\mu\nu\rho}$ as \begin{equation}
h^+_{\mu\nu\rho} = \frac 14 \left(H_{\mu\nu\rho} - 3 u_{[\mu}\cdot\left(2\ft{H^-_{\nu\rho]}}{\sqrt{u^2}}\right)\right)\, .
\end{equation}
The last term can be seen as the product of  $u_\m$, the gauge field of the second symmetry as argued before \eqn{gauge}, and \eqn{symm2}, the $II$-transformation of $B_{\m\n}$. We define covariant derivatives of a field as its partial derivative minus the gauge transformation(s) where the parameter is replaced by the gauge field. In this sense, $h^+_{\m\n\r}$ can be considered as the fully covariant field strength of $B_{\m\n}$, where `fully' means {\it also with respect to gauge symmetry} $II$.  $h^+_{\m\n\r}$ is automatically self-dual in this way. The new supersymmetry transformation rule, which gives rise to the old one by imposing \eqn{selfdual}, is then:
\be\label{susy2psi}
\d \psi^i = \frac{1}{12} h^{+}_{\m\n\r} \c^{\m\n\r} \e^i
+\frac{1}{4}\not\!\! \cD \phi^{ij}
\e_j - \phi^{ij}\eta_j\, .
\ee
By considering $h^+_{\m\n\r}$ as the fully covariant field strength of $B_{\m\n}$, we understand better why this term appears in the supersymmetry transformation rules. Before, it was unclear where this self-dual tensor came form. 
The supersymmetry transformation rule depends also on the matter terms, proportional to $\phi_{ij} T^{ij}_{abc}$ in $h^+_{\m\n\r}$. This is necessary to realize the algebra (\ref{QQ}) (and later also to find the invariance of the action). This was one of the reasons to include also the matter term in \eqn{covH}.\\
The covariant derivative of $\psi^i$ and $\phi^{ij}$ and the covariant d'Alembertian of $\phi^{ij}$ are:
\begin{eqnarray}
{\cal D}_\mu \psi^i &=& \left(\partial_\mu - \ft 52 b_\mu +
\ft 14 \o_\mu{}^{ab}\gamma_{ab}\right )\psi^i -\ft 12 V_\mu^i{}_j\psi^j\nn \\
&&-\ft 1{12} h^+_{abc}\gamma^{abc}\psi_\mu^i - \ft 14\left(
\not\!\! {\cal D}\phi^{ij}\right)
\psi_{\mu j} + \phi^{ij}\phi_{\mu j}\ ,\\
{\cal D}_\mu \phi^{ij} &=& \left (\partial\mu -2b_\mu\right )
\phi^{ij} + V_\mu{}^{[i}{}_k \phi^{j]k} + 4 \left (
{\bar\psi}_\mu^{[i} \psi^{j]} - {\rm trace}\right )\ , \\
\label{box}
{\cal D}^a{\cal D}_a \phi^{ij} &=& \partial^a{\cal D}_a \phi^{ij}
-3b^a {\cal D}_a\phi^{ij} - V_a^{[i}{}_k {\cal D}^a\phi^{j]k} +\o_a{}^{ab}
{\cal D}_b\phi^{ij} -4 f_a{}^a\phi^{ij}\nn \\
&&+4 {\bar \psi}_a^{[i}{\cal D}^a \psi^{j]} - \ft 2{15} \left(
{\bar\psi}^a_l\gamma_a\chi^{(i,k)l}\phi^j {}_k - i\leftrightarrow j\right) \\ \nn
&&- \ft 16 {\bar\psi}^a_k\gamma_a \gamma^{bcd}T^{k[i}_{bcd}
\psi^{j]}
-4 {\bar\phi}_a^{[i}\gamma^a \psi^{j]} - ({\rm trace})\, ,
\end{eqnarray}
where $f_a{}^a = e_a{}^\m f_\m{}^a$ and the trace guarantees that the contraction with $\Omega_{ij}$ vanishes.
%\subsubsection{The realization of the algebra}
\par The algebra of $Q$- and $S$-transformations contains field-dependent transformations. When using (\ref{susyBmn}), (\ref{susy2psi}) and (\ref{susyphi}), the transformation rules for the self-dual tensor multiplet and the transformation rules for the Weyl multiplet, the commutators of the supersymmetries become
\bea
\left[\d_Q(\vare_1), \d_Q(\vare_2)\right] & = & 
\ft 12 ({\bar \vare}^i_2\c^\m\vare_{1i}) \, \hat\cD_\m 
+ \d_M(-\ft 12 {\bar \vare}_2^iT_{ij}^{abc}\c_c\vare_1^j) \nn\\
&&+\d_S(-\ft2{45}{\bar \vare}_{1[k}\gamma_a\vare_{2j]}\c^a\chi^{(i,j)k}) 
+ \d_K(-\ft 18 {\bar \vare}_2^i\c_b\vare^j_1\, {\cal D}_cT^{abc}_{ij}) \nn\\
&&+ \d_I (- \ft 12{\bar \vare}^i_2\c_\m\vare_1^j\phi_{ij}) \label{QQ}\, ,\\ 
\left[\d_S(\eta), \d_Q(\vare)\right] & = & \d_D(-\ft 12 {\bar \vare}^i\eta_i) + \d_M(-\ft 12 {\bar \vare}^i\c^{ab}\eta_i) +\d_{USp(4)}(-4{\bar \vare}^{(i}\eta^{j)}) \label{QS} \, , \\
\left[\d_S(\eta_1), \d_S(\eta_2)\right] & = & \d_K(-\ft 12{\bar\eta}^i_2\c^a\eta_{1i})\label{SS}\, .
\eea
Some comments can be made about this algebra:
\par The $\hat\cD_\m $ in the first term of (\ref{QQ}) is a covariant general coordinate transformation \cite{sctc}. This includes all the superconformal transformations, gauge symmetry $I$ and {\it also gauge symmetry} $II$. The covariant general coordinate transformation with derivative $\hat\cD_\m $ contains terms which are a gauge transformation $I$, $II$ or $III$.  There is no natural gauge field for symmetry $III$, so $\hat\cD_\m $ is only covariant with respect to symmetry $I$ and $II$. 
\par The term in the algebra with gauge transformation $I$ was already introduced in \cite{BSVP1} for the case of the $(1,0)$ self-dual tensor. This means that each of the gauge symmetries plays a specific role in the algebra: symmetry $I$, $II$ and $III$ in the covariant general coordinate transformations, while symmetry $I$ appears also with a field-dependent gauge transformation.
\par With $u_\m$ as the gauge field for symmetry $II$, the algebra on $a$ becomes 
\bea
\left[\d_Q(\vare_1), \d_Q(\vare_2)\right] a & = & 
\ft 12 ({\bar \vare}^i_2\c^\m\vare_{1i}) \hat\cD_\m a \nn \\
& = & \ft 12 ({\bar \vare}^i_2\c^\m\vare_{1i}) (\partial_\m a - u_\m) \nn \\
& = & 0\, ,
\eea
both for the rigid and the local supersymmetric case.
So, $a$ is a supersymmetric singlet \cite{bkr}. 
\par Using the prescription of appendix $B$ in \cite{BSVP}, it is possible to re-derive the $(1,0)$-algebra. The term with the Lorentz rotation and gauge symmetry $I$ give immediately the right term. The terms with special conformal symmetry and special supersymmetry need more work. The relevant terms of the constraints and transformation rules differ slightly for the different cases. Also the shifts in $\phi_\m^i$ and $f_\m{}^a$ when going to $(1,0)$ need to be taken into account. Also in simple chiral supergravity, symmetries $I$, $II$ and $III$ appear in the commutator of two supersymmetries.

\par Finally, we remark that the algebra is not realised off-shell. In the transformation of $H^-_{\m\n\r}$ (\ref{trsusy}), the equation of motion of the spinor still appears. Therefore, in the commutator of two supersymmetries on $\psi^i$ there also appears an equation of motion. Also the commutator of symmetry $II$ with supersymmetry contains terms with equations of motion.

\section{The superconformal action}
\label{s:action}

In \cite{ckvp} was already proven that the action of the rigid model has  superconformal invariance. Here, we give the action for  the self-dual tensor multiplet with local $(2,0)$ superconformal symmetry. We also see that the action, as expected, gives rise to the equations of motion of \cite{BSVP}.

%\subsection{The invariance of the action}

\par The following action is invariant under local superconformal transformations and under the 3 bosonic symmetries (\ref{symm1}), (\ref{symm2}) and (\ref{symm3}):
\bea\label{action}
S & = &\int d^6 x \sqrt{g} \left[ - H^-_{\m\n} H^{*\m\n} -\ft 16 H^{*\m\n\r} \left(3\bar\psi_{[\mu}^i\gamma_{\nu\rho]}\psi_i -\ft 32 \bar \psi_{[\mu}^i\gamma_\nu\psi_{\rho ]}^j\phi_{ij} 
- \ft 12 \phi_{ij}T^{ij}_{\mu\nu\rho}\right)
\right. \nonumber \\
&&-4\bar\psi^i \cDs' \psi_i -\bar\psi^i_\m\gamma^\m \cDs' \psi^j\phi_{ij} \nonumber \\
&&+\ft 14 \phi^{ij} \left( \partial^a {\cal D}_a\phi_{ij} 
-3b^a{\cal D}_a\phi_{ij} +V^a_{ki}{\cal D}_a\phi_j{}^k +\omega_a{}^{ab}{\cal D}_b\phi_{ij} -4f_a{}^a\phi_{ij} \right. \nonumber \\
&& \left. + 4\bar\psi_{ai}\cD'^a \psi_j +4\bar\phi_{ai}\g^a\psi_j \right) \nonumber \\
&&-\ft{8}{15}\bar\psi^i\chi_i^{kl}\phi_{kl} +\ft {1}{60}D^{ij,kl}\phi_{ij}\phi_{kl} +\ft 13 \bar\psi_iT^{ij}\cdot\g\psi_j \nonumber \\
&&+\ft{1}{15} \bar\chi^i_{kl}\phi^{kl}\phi_{ij}\g^\m\psi_\m^j 
+\ft {1}{24}\bar\psi_\m^i\g^\m \phi_{ij} T^{jk}\cdot\g\psi_k 
-\ft {1}{48}\bar\psi^i_\m\g^\m \phi_{jk}T^{jk}\cdot\g \psi_i\nonumber \\
&&-\ft {1}{288} \bar\psi^i_\m\g^\m T^{jk}\cdot \g\g^\n\psi_\n^l\phi_{ij}\phi_{kl} 
+\ft {1}{192} \bar\psi_\m^i\g^\m T^{jk}\cdot \g\g^\n\psi_\n^l\phi_{jk}\phi_{il} \nonumber\\
&&\left. -\ft{1}{720}\bar\psi_\m^i\g^\m T_{ij}\cdot \g\g^\n\psi_\n^j\phi_{kl}\phi^{kl} 
%\bar\psi_\m^i\g^\m T^{jk}\cdot\g\g^\n\psi_{\n i}\phi_j{}^l\phi_{kl}
\right]\, .
\eea
In (\ref{action}) appears the covariant derivative of $\psi^i$ without the $h^+_{\m\n\r}$-term:
\bea
\cD'_\m \psi^i & \equiv & 
\left(\partial_\m - \ft 52 b_\m +
\ft 14 \o_\m{}^{ab}\c_{ab}\right)\psi^i -\ft 12 V_\m^i{}_j\psi^j\nn
\\
&&- \ft 14 \left( \not\!\! \cD\phi^{ij}\right)
\psi_{\m j} + \phi^{ij}\phi_{\m j}\, .
\eea
Further, $T_{ij}\cdot\c = T_{ij}^{abc}\c_{abc}$.\\
This action describes the coupling of a self-dual tensor in six dimensions to a conformal supergravity background. 
The first gauge symmetry imposes that $B_{\m\n}$ only appears in a field strength in the action. The second and the third gauge symmetries impose the form of the action as given in (\ref{action}) for terms that transform with respect to one of these symmetries. 
The second part of the first line contains only covariantization terms. It is imposed by the third gauge symmetry and can also be found in the local super-Poincar\'e actions for self-dual tensors \cite{DLT}. It is absent for free chiral bosons. All the coefficients of the action are fixed by imposing invariance under supersymmetry and special supersymmetry.

%\subsection{The field equations}

\par The action (\ref{action}) gives rise to the following field equations for $B_{\m\n}$, $a$, $\psi_i$ and $\phi^{ij}$:
\bea\label{eomBmn}
{\cal G}^{\m\n} & = & \partial_\rho\left( 4eh^{+\m\n\r} -\ft 16\varepsilon^{\m\n\r\s\tau\phi}C_{\s\tau\phi}\right) = 0\, , \\
\label{eoma}
{\cal A} & = & \partial_\phi\left(\frac1{\sqrt{u^2}}
\epsilon^{\mu\nu\rho\sigma\tau\phi} H^-_{\mu\nu}
H^-_{\rho\sigma}v_\tau  \right) = 0 \, ,\\
\label{eompsi}
\Gamma^i & = & \not\!\! \cD\psi^i -{\ts{1\over 15}}\phi^{kl}}\chi^i_{kl} -{\ts{1\over 12} T^{ij}_{abc}\c^{abc}\psi_j = 0 \, , \\
\label{eomphi}
C_{ij} & = & \cD^a \cD_a \phi_{ij} -{\ts{1\over
15}}D_{ij}^{kl}\phi_{kl} +{\ts{1\over 3}} h^+_{abc} T_{ij}^{abc} +
{\ts{16\over 15 }}{\bar \chi}_{ij}^k\psi_k = 0\, .
\eea
The equations (\ref{eompsi}) and (\ref{eomphi}) are the equations of motion as derived in \cite{BSVP}. Also  the self-duality condition (\ref{selfdual}) corresponds corresponds to the self-duality equation derived there. The only difference is the one but last term in (\ref{eomphi}), but that is a term proportional to the self-duality condition. Rewriting (\ref{eomBmn}) gives:
\be
{\cal G}^{\m\n} = -\vare^{\m\n\r\s\tau\phi}\partial_\rho\left(v_\s H^-_{\tau\phi}\right) = 0\, .
\ee
Analogous to the rigid case, the self-duality condition (\ref{selfdual}) can be found by a gauge-choice of symmetry $III$ for the most general solution of this equation of motion (if there are no global obstructions). So, this action describes a self-dual two-form. 
\par The description using this action is more satisfactory than the one of \cite{Ric}.  There, first the action for an ordinary tensor is written down and the equation of motion is derived. Only then, the self-duality condition is imposed. Here the self-duality is automatically incorporated in the action and follows from the equation of motion.

\par The equations of motion transform in the following way into each other under (special) supersymmetry:
\begin{eqnarray}\label{trsusy}
\delta H^-_{abc} & = & -\ft 12{\bar\epsilon}^i\gamma_{abc}
\Gamma_i\, ,\nn \\
\delta{\cal A} & = & -  \partial_\phi\left(\frac1{\sqrt{u^2}}
\epsilon^{\mu\nu\rho\sigma\tau\phi} H^-_{\mu\nu}
{\bar \epsilon}^i \gamma_{\rho\sigma\upsilon}\Gamma_i v^\upsilon v_\tau  \right) \nn\\
\delta {\cal G}_{ab} &=& - {\bar\epsilon}^i\gamma_{abc}
 {\cal D}^c\Gamma_i\, ,\nn \\
\delta \Gamma^i &=& \ft14 C^{ij}\epsilon_j 
+ \ft18 \gamma^{ab}\epsilon^i {\cal G}_{ab}
-\ft18 \gamma^\mu\gamma_a\Gamma^i {\bar\epsilon}^i
\gamma^a\psi_{\mu i}\, ,\\
\delta C^{ij} &=& -4{\bar\epsilon}^{[i}
\not\!\!  {\cal D}
\Gamma^{j]} + \ft18 {\bar\psi}^{[i}_\mu\gamma_a\Gamma^{j]}
{\bar\epsilon}^i\gamma^a\psi^{\mu i} -8{\bar\eta}^{[i}\Gamma^{j]} -
({\rm trace})\, .\nonumber
\end{eqnarray}

\section{Conclusions}

In this article we have interpreted the second bosonic symmetry as a gauge symmetry with gauge field $u_\m$. Its role in the supersymmetry transformation rules and in the realization of the $OSp(8^*|4)$-algebra has become more clear. It appears on the right hand side in the covariant general coordinate transformation in the commutator of two supersymmetries. One also finds that the auxiliary scalar $a$ is a supersymmetric singlet. Gauge symmetry $I$ appears as a field-dependent gauge transformation in the commutator of two supersymmetries.

\par Furthermore, we have given the action for a self-dual tensor multiplet coupled to a chiral conformal gravity background with $(2,0)$ supersymmetry. This action gives rise to the equations of motion found earlier in \cite{BSVP}. The self-duality follows from the action by imposing a gauge fixing of symmetry $III$.

\par The action, \eqn{action}, is the most general one for the coupling of self-dual tensor multiplets to (conformal) supergravity. Using the prescription of \cite{BSVP}, the Poincar\'e action for $n$ self-dual tensor multiplets can be derived from this superconformal formulation. One starts with $n+5$ tensor multiplets in the vector representation of $SO(n,5)$. Imposing the suitable constraints that gauge-fix the appropriate superconformal symmetries (dilatations, special conformal symmetry and special supersymmetry) yields the coupling of $n$ self-dual tensor multiplets to $(2,0)$ Poincar\'e supergravity. The $5n$ scalars of these $n$ tensor multiplets parameterize the coset $\frac{SO(n,5)}{SO(n)\times SO(5)}$ \cite{romans,BSVP}.

\par Using the procedure in \cite{BSVP} to move from $(2,0)$ to $(1,0)$, one discovers the action for a self-dual tensor multiplet in a $(1,0)$ superconformal gravity background. For this case, a similar procedure is possible to move further towards a Poincar\'e description. Starting from $n+1$ tensor multiplets in the vector representation of $SO(n,1)$ and imposing the right constraints will break the superconformal symmetry. This should give the actions of \cite{DLT}. The scalars in the Poincar\'e theory are in $\frac{SO(n,1)}{SO(n)}$ \cite{romans}. In this context, with $(1,0)$ supersymmetry, couplings of actions for self-dual tensor multiplets to other matter multiplets are possible. 

\par An interesting project would be to calculate the conformal anomaly for the self-dual tensor, e.g. by following the approach in \cite{hennsken} and using the ghost sector of the action for the self-dual tensor of \cite{vdbvh}. This weak-coupling calculation can be compared with the strong-coupling result in \cite{grahamwitten}.

\section*{Acknowledgments}

\noindent
I thank Toine Van Proeyen and Piet Claus for many stimulating discussions and useful comments and Kostas Skenderis for suggesting the project in the last paragraph. This work was supported by the European Commission TMR programme ERBFMRX-CT96-0045.

\end{document}